\documentclass{WileyMSP-template}
\usepackage{amsmath}
\usepackage[numbers]{natbib}
\begin{document}

\pagestyle{fancy}
\rhead{
}

\title{In-Liquido Computation with Electrochemical Transistors and Mixed Conductors for Intelligent Bioelectronics}

\maketitle


\author{Matteo Cucchi}
\author{Daniela Parker}
\author{Eleni Stavrinidou}
\author{Paschalis Gkoupidenis}
\author{Hans Kleemann}


\dedication{}

\begin{affiliations}
Dr. M. Cucchi $^{1,2,*}$\\
$^1$ Ecole Polytechnique Fédérale de Lausanne (EPFL), Laboratory for Soft Bioelectronic Interfaces, Neuro-X Institute, Geneva, Switzerland\\
$^2$ Dresden Integrated Center for Applied Photophysics and Photonic Materials (IAPP), Technische Universität Dresden, Dresden, Germany\\

D. Parker and Dr. E. Stavrinidou\\
$^3$ Laboratory of Organic Electronics, Department of Science and Technology, Linköping University, SE-60174, Norrköping, Sweden\\
Dr. H. Kleemann\\
$^2$Dresden Integrated Center for Applied Photophysics and Photonic Materials (IAPP), Technische Universität Dresden, Dresden, Germany

Dr. P. Gkoupidenis $^3$\\
$^4$ Max Planck Institute for Polymer Research, Mainz, Germany\\

\end{affiliations}


\keywords{Neuromorphic computing, Electrochemical transistors, Bioelectronics, In-liquido computation, Intelligent bioelectronics}

\begin{abstract}
Next-generation implantable computational devices require long-term stable electronic components capable of operating in, and interacting with, electrolytic surroundings without being damaged. 
Organic electrochemical transistors (OECTs) emerged as fitting candidates. However, while single devices feature impressive figures of merit, integrated circuits (ICs) immersed in a common electrolytes are hard to realize using electrochemical transistors, and there is no clear path forward for optimal top-down circuit design and high-density integration.
The simple observation that two OECTs immersed in the same electrolytic medium will inevitably interact hampers their implementation in complex circuitry. The electrolyte's ionic conductivity connects all the devices in the liquid, producing unwanted and often unforeseeable dynamics. Minimizing or harnessing this crosstalk has been the focus of very recent studies.
 In this Perspective, we discuss the main challenges, trends, and opportunities for realizing OECT-based circuitry in a liquid environment that could circumnavigate the hard limits of engineering and human physiology. We  analyze the most successful approaches in autonomous bioelectronics and information processing. Elaborating on the strategies to circumvent and harness device crosstalk proves that platforms capable of complex computation and even machine learning can be realized in-liquido using mixed ionic-electronic conductors.
\end{abstract}

\section{Introduction}
With the advent of solid-state transistors in the 1950s, digital computers took off. Since then, nanoelectronics and information processing have become arguably the most successful technological endeavors ever pursued in human history. Their incredible advances stem from a sophisticated and precise top-down integration that ranges from user-friendly interfaces down to nanometric transistors. 
Electrodynamics is exploited to achieve Boolean logic, allowing fast, precise, general-purpose, and handy data processing. This paradigm is excellent for number crunching and logic operations.
Furthermore, in the last two decades, the domain of competence of digital electronics extended to tasks such as detection, recognition, and classification of patterns thanks to the advent of machine learning (ML) and artificial intelligence (AI). The success of digital architecture is further emphasized by the fact that integration, circuit design, and materials can be applied seamlessly to a wide range of applications, from smartphones and robotics to the automotive industry and space exploration. 
However, exploiting the conventional digital computation scheme for some specific applications is still challenging. For instance, coupling electronics to biological systems remains a thorny task. Nonetheless, the field is very active, and a tremendous effort is being dedicated to this goal, pushed by the undisputed potential of integrating electronic circuits within the human body. Life-saving examples, such as the pacemaker, the insulin pump, and the cochlea implant, are already widely employed in clinical practice. At the same time, it is interesting to observe that current bioelectronics typically perform a "passive" function,  eliciting signals in the presence (or absence) of a met condition $e.g.$, hyperglycemia or arrhythmic heartbeat, or recording biopotentials (Fig. 1a). Instead, complex computation beyond the mere sensing is difficult and application-specific ICs are generally employed which must be encapsulated and placed nearby the sensing unit  \cite{yoo2021neural,shin2022256}, but the  energy-intensive computation and pattern recognition that can be carried out with modern ML techniques cannot be yet implemented in implanted devices. The reasons are multiple and straightforward: firstly, the human body strongly hampers the design and fabrication of implantable electronics because of the strict limitations in terms of heat dissipation, material stability, size, and biological compatibility. Depending on the application and on the anatomical target, such limitations may range from moderate to deal-breakers. Notable examples of electrical  \cite{piech2020wireless} and optical  \cite{kathe2022wireless} stimulation on in-vivo freely-behaving animal models confirm both the cumbersome electronics needed for autonomous functioning of closed-loop bioelectronics and, at the same time, its huge medical potential for humans. 
Moreover, complex digital computation requires a significant energy supply, a problem typically worked-around (e.g., in smartphone) using cloud-computing. However, when dealing with sensitive biometric data, privacy issues immediately arise. On-chip (edge) computation represents a suitable alternative. Therefore, closed-loop systems and edge computation are to be preferred  \cite{berggren2020roadmap}. Because of these impediments, the employment of ML on in-vivo models has only been exploited primarily offline $i.e.$, after the patient's analysis, using a conventional computer, or limited to simple in-vitro models.  \\
Finally, electronics must interface the biological electrical circuitry: the two have completely different blueprints, with the latter relying on a complex system of chemical reactions and feedback loops, a system often referred to as "wetware". The wetware is capable of processing chemo-electrical signals and respond in a way that counteract on malign patterns (Fig. 1a). Our biochemical machinery is source of inspiration for better medical treatments that combines low energy needs and compatibility with our body. In this framework, the design of autonomous intelligent bioelectronics that can sense external stimuli, learn how to classify them, and act on them is highly sough-after (Fig. 1a).\\
With these challenges in mind, new advances in the field of computer science, physics and engineering may help finding novel paradigms that allows energy-efficient and low-latency computation using systems that could better interface the biological environment. Recently, there has been increasing interest in what is often called "in-materio" computation: the idea is to exploit the dynamical properties of a material -- or a generic physical system-- to carry out information processing directly "on-chip" exploiting the response of such system to external stimuli  \cite{Tanaka2019, cucchi2022hands}. The "analog" nature of the environment-chip interaction ensure fast and energy-efficient signal communication, without the need of a analog-to-digital conversion. Such a paradigm ensures low power consumption and miniaturized chips that can be seamlessly implanted in living organisms or exploited for in-vivo analysis.
 \begin{figure}
    \includegraphics[width=\linewidth]{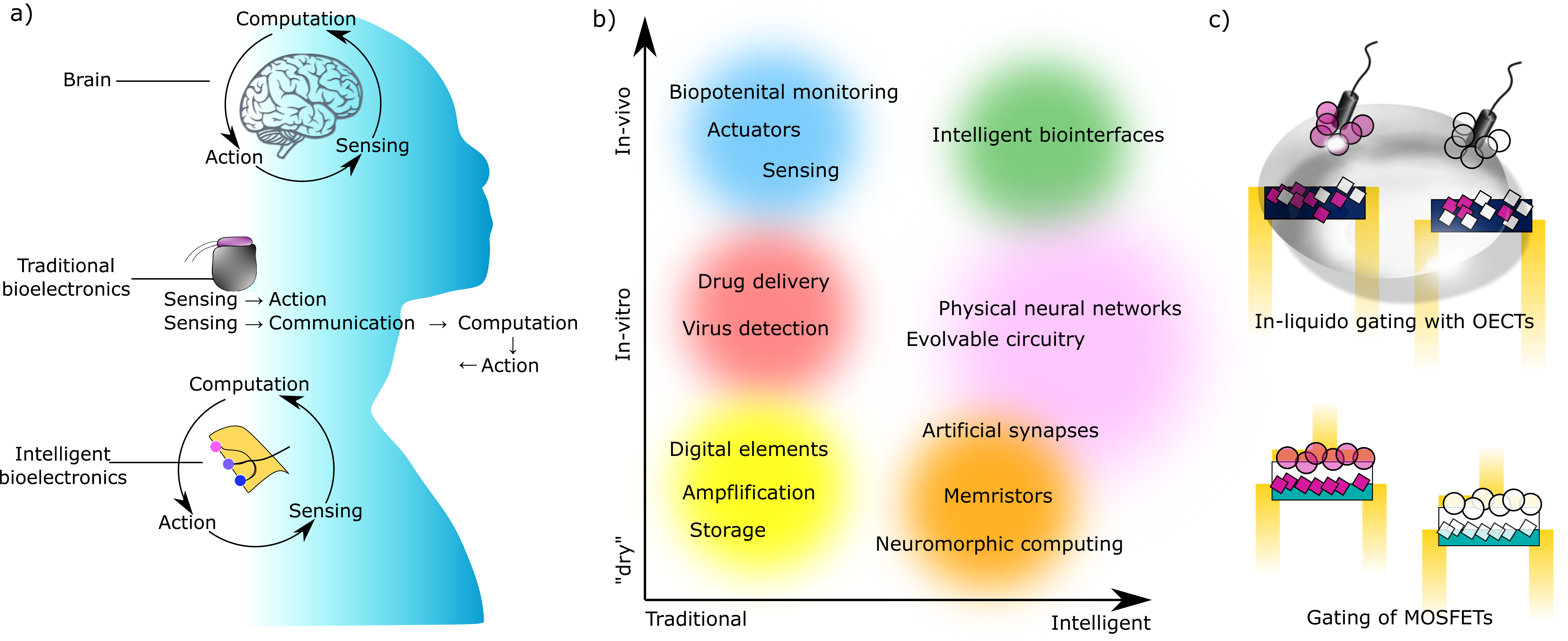}
    \caption{a) Sketch of the traditional working mechanisms of today's medical implants compared to the ideal smart implants. b) Overview of the applications that OECTs have been demonstrated for, organized by the level of biological implementation and adaptive complexity.
    c) Comparison between gating MOSFETs in traditional electronics, where each channel is only influenced by its "parent" gate and OECTs, where each gate will affect every channel immersed in the same medium (square and circles indicate positive and negative charges, while pink and white denote the charges coming from one transistor or the other).}
 \end{figure}
 
Embracing in-materio computing means supplanting the digital paradigm in favor of intrinsic physical properties. How these are exploited for a specific computational task may be material- or application-specific. Among these systems, materials that are able to carry out information processing in electrolytic environments could be leveraged for "in-liquido" computation i.e. perform pattern recognition and inference in liquid environments.\\
In this scenario, polymers capable of conducting both ions and electrons, the so-called organic mixed ionic-electronic conductors (OMIECs) emerge as prime candidates. OMIECs are gated through the electrolyte, hence forming an electrochemical transistor (OECT). OECTs have recently been in the spotlight because of their peculiar physical and chemical properties, and they have been shown to be suitable for an impressively wide plethora of applications, from conventional logic circuitry to brain-inspired and analog computation. Moreover, clever device design and fabrication allow their employment in different environments, from "dry" chips to in-vitro and in-vivo  \cite{owens2010organic,khodagholy2015neurogrid,khodagholy2013vivo} (Fig. 1b).\\
In this perspective, we analyze the recent efforts that deal with the operation of OECT-based circuitry in liquid environments. We explain why digital electronics is hard to achieve even for simplistic circuitry because of the difficult task of avoiding device-to-device crosstalk: indeed, while in traditional circuits there is a defined and localized set of inputs and outputs per device, all electrochemical transistors communicate through the electrolyte. The conductivity of the electrolyte is not high enough to cause a short-circuit, but mediates a strong capacitive coupling between all devices (Fig. 1c). Very often, the IV characteristics of single electrochemical transistors are praised and envisioned to be of great utility for bioelectronic applications where necessarily "homeostatic conditions" apply, i.e. a common electrolyte interacting with the devices, without foreseeing such crosstalk. Only very recently, researchers started to explore how to circumnavigate and even harness this crosstalk, with promising demonstrations for emerging intelligent bioelectronics. The results are encouraging although so far limited to prototypical demonstrations: a clear path forward for real-life and clinical applications is yet to be identified.
Our analysis focuses on OECTs and OMIECs-based devices, but general principles may be applied to other technologies operating in global electrolytic conditions e.g. Ion-sensitive FETs, electrolyte-gated FETs, and electrochemical transistors based on graphene and graphene oxide. As it turns out, if one wishes to have such circuitry in contact with the electrolytic environment, building logic gates and other basic components, cornerstone of the digital architecture, becomes extremely challenging. However, the digital paradigm can be left behind and the peculiar nonlinear properties of electrochemical transistors and the complex spatio-temporal coupling among devices can be leveraged for computation. In particular, the extended transients given by the different dynamics of ions and electrons can be exploited to operate unconventional computing schemes such as brain-inspired and random neural networks and to realize implantable computational platforms capable of pattern recognition and inference in biological environments.

\section{In-Liquido Operation of OECTs}

OMIECs are typically  $\pi-$conjugated polymers with low oxidation or reduction potential (see e.g., review article by Paulsen \textit{et al.}  \cite{paulsen2020organic}). 
The rise of organic mixed conductors for in-liquido electronics stems from a unique mechanism that couples the electronic and ionic currents flowing through the polymeric matrix. To illustrate this mechanism, as well as other examples throughout this Perspective, we 
use poly(3,4-ethylenedioxythiophene) polystyrene sulfonate (PEDOT:PSS) as model, a p-type OMIEC that produces normally-on, depletion-mode OECTs. While historically PEDOT:PSS is by far the most employed OMIEC, recently, many materials were shown to transport ions and electrons, including n-type polymers. 
For a detailed account of the physical and chemical properties of OMIECs and OECTs, the reader can refer to recent work such as Refs.  \cite{paulsen2020organic,cucchi2022thermodynamics}.

\begin{figure}
	\includegraphics[width=\linewidth]{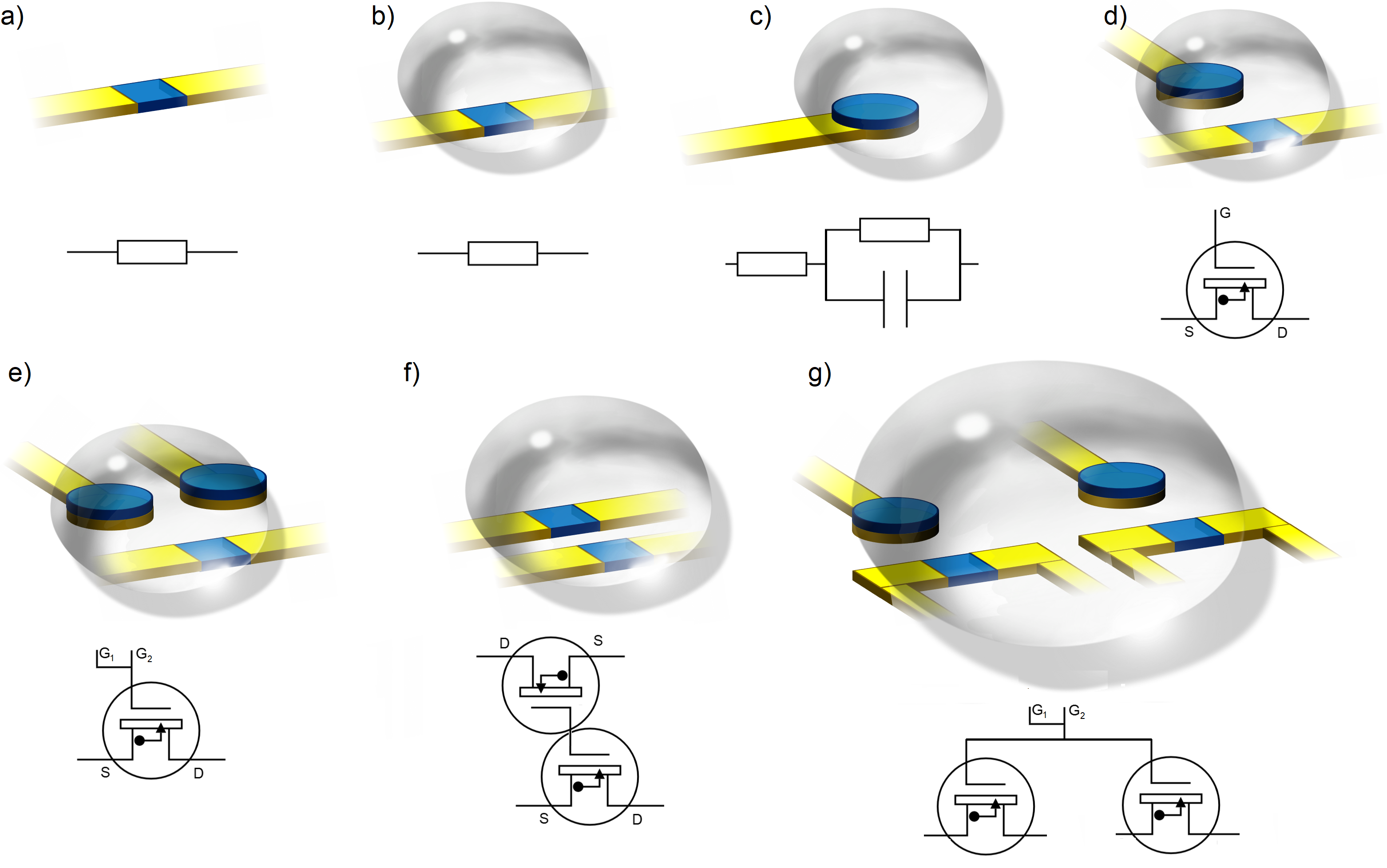}
	\caption{Sketch of the basic OMIECs-based device architectures used throughout this work to describe the electrical properties of single and multi-device systems and crosstalk: a) a single OMIEC channel in the dry state behaves as a resistor, b) a single OMIEC channel in electrolyte also behaves like a resistor, c) an OMIEC/electrolyte interface (described by a Randles circuit), d) an OECT, e) an OECT with two gates, f) two channels without "proper" gate, and g) two neighboring OECTs.}
	\label{fig:panel2}
\end{figure}

In a simplified picture, in its pristine (dry) state (prior the electrolyte immersion), each negative PSS$^-$ moiety is bound to monomeric unit of PEDOT; the latter features a very low oxidation potential, hence it undergoes oxidation, breaks one of the double bonds on the $\pi$-conjugated system and generates a free hole (or more accurately a polaron). Given the good charge carrier mobility granted by the  $\pi$-conjugated backbone, as well as the weak electrostatic attraction exerted by the PSS$^-$, the hole is free to drift under the influence of an external electric field. The electronic conductivity of the film $G$ is
\begin{equation}
	G=\mu \frac{Wt}{L} e\rho_0 = G_0.
	\label{eq:conductivity}
\end{equation}
where $\rho_0$ is the concentration of the dopants PSS$^-$ which generate a free charge, and $W$, $t$, and $L$ are the geometrical width, thickness and length of the film, respectively. In their dry state, OMIECs  behave like resistors (Fig. \ref{fig:panel2}a). The interesting properties of OMIECs arise when they are immersed in an electrolytic solution. It is important to notice that OMIECs swells significantly when placed in a suitable solvent and the polyelectrolyte (here PSS) mediates the ionic conduction through the polymeric matrix.  This makes OMIECs ionic conductors  \cite{stavrinidou2013direct}, therefore granting a unique and simultaneous electronic and ionic conduction that can be readily harnessed to interface biological and artificial systems.
The ions coming from electrolyte are able to counterbalance the charged moiety of the polyelectrolyte, therefore lowering the number of polarons and so the polymers conductance.  As an example, for PEDOT:PSS immersed in an aqueous solution of NaCl, such process obeys the following chemical reaction
\begin{equation}
	\text{PEDOT$^+$:PSS$^-$} +  \text{Na}^+  + \text{e}^-  \rightleftharpoons \text{PEDOT}^0 + \text{Na$^+$:PSS$^-$}.
	\label{eq:reaction}
\end{equation} 
As each cation penetrating the film (ionic current) directly affects the hole concentration (and therefore the electronic current), ionic current and electronic current are intimately coupled in mixed conductors. The transformation of polarons (PEDOT$^+$) in favor of reduced PEDOT$^0$ can be regarded as an electrochemical dedoping. The process described in Eq. \ref{eq:reaction} establishes an equilibrium in which a fraction $\phi$ of the total PEDOT$^+$ amount ($\rho_0$) is reduced to PEDOT$^0$.
Therefore, when immersed in an electrolytic solution, the charge carrier concentration $\rho$ of PEDOT  differs from the dry state, and Eq. \ref{eq:conductivity} must be corrected as 
\begin{equation}
	G=\mu \frac{Wt}{L} e\rho=\mu \frac{Wt}{L} e\rho_0(1-\phi) =G_0 (1-\phi).
	\label{eq:conductivity2}
\end{equation}
 Romele \textit{et al.} showed that the salt concentration of the electrolyte does not affect the quantity $\phi$, which is therefore a material-specific property that can be explained in terms of entropy of mixing  \cite{romele2019ion,cucchi2022thermodynamics}. The existence of this equilibrium has profound consequences on the device physics inasmuch the $non-gated$ channel is in an intermediate state between the fully on and fully off state, depending on the value of $\phi$. Therefore, the gate perturbs this equilibrium both with a postive and negative bias, which makes the OMIEC responsive to a larger plethora of signals rather than the ones of one polarity only.   When the film interacts with the electrolyte its electrical properties change and the OMIEC typically behaves like a resistor (with a resistance different from the "dry" scenario, Fig. \ref{fig:panel2}b).\\
 
 The ease with which such polymers reversibly undergo reduction and oxidation grants them exceptional properties as solid/liquid electrical interfaces, especially considering the minimal contact resistance between OMIECs and noble metals. Therefore, by generating and annihilating free charges, OMIECs are capable of transducing efficiently electronic currents into ionic currents, a sought-after feature for bioelectronic biotic/abiotic interfaces. This mechanism stands out when compared to typical low-impedance yet unstable and sometimes toxic non-polarizable electrodes (e.g., Ag/AgCl), as well as to durable yet highly resistive polarizable electrodes (Pt, Au, etc) (Fig. \ref{fig:panel2}c). 

The good ion-electron coupling can be better leveraged by employing a third electrode in order to  "gate" an OMIEC thin film (channel) as displayed in Fig. \ref{fig:panel2}d, thus forming an organic electrochemical transistor.
In this way, the passive diffusive process of the ions penetrating the film can be actively controlled and steered from equilibrium by applying a potential difference between the gate electrode and the channel. The reader can refer to detailed articles on OECTs operation and characterization  \cite{romele2019ion, Berggren2022}.

In the following, we simplify the treatment and mathematical modeling of OECTs using the traditional description of OECTs as thin-film transistors, an approach that despite strong assumptions turns out to approximate well the linear behavior of such devices  \cite{bernards2007steady}. Within this Bernards model, the drain current $I_d$ varies in  presence of a potential difference between source and gate $V_g$ and between source and drain $V_d$ as

\begin{equation}
I_d =G_0 (1- \frac{V_g - \frac{V_d}{2}}{V_p})V_d,
\label{eq:oect}
\end{equation}  
where $V_p$ is the pinch-off voltage and it is related to the film's volumetric capacitance $C^*$, according to $V_p=e\rho_0/C^*$. \\
In this Perspective, Eq. \ref{eq:oect} is used to model OECTs, and it is then extended in Sec. \ref{sec:globalgating} to multi-device operation to model device-to-device coupling.

Differently from the two-electrode case, in an OECT, the electrolyte concentration affects the transfer characteristics of the transistor, due to the voltage drop across the gate-electrolyte-channel subcircuit, which typically follow a Nernstian dependence. The voltage drop at the gate-electrolyte interface is further influenced by the gate's material and size, an effect that has been thoroughly exploited to demonstrate OECTs as ion sensors  \cite{lin2010ion,tseng2021membrane}. On this basis, only a fraction of the applied gate voltage is "felt" by the channel. Such fraction is often referred to as "gate coupling" and its maximization is key for high-performance OECTs. Moreover, the gate's electrode can be chosen to trigger Faradaic reactions that allow sensing of biologically-relevant neutral chemicals, such as glucose and hydrogen peroxide  \cite{yan2020high}.

The ionic circuit is generally modeled with a Randles cell for the gate-electrolyte interface, with a resistor for the bulk electrolyte (cf. Figure \ref{fig:panel2}c), and another Randles cell for the electrolyte-channel interface.
Typically, at low frequencies, the interfaces dominate the impedance of the ionic circuit.
OECTs can attain impressive transistor performances: they can be switched off with less than 1 V, have high transconductance, and feature a sub-threshold slope that approaches the thermodynamic limit  \cite{wang2022realizing,khodagholy2013high,nishinaka2021high,weissbach2022photopatternable}. Note that these characteristics can be leveraged for autonomous and self-powered electronics, key for battery-free implantable electronics. Additionally, by engineering a vertical channel, impressive current densities have been shown  \cite{lenz2019vertical}, and by minimizing the distance between channel and gate, the response time can be as low as 31 $\mu$s  \cite{spyropoulos2019internal}. Moreover, the basic architecture of the OECT can be extended and modified in order to have different devices such as diodes, memristors, batteries, etc \cite{shen2021device}.
A single OECT-based ion sensor can easily be employed for in-vivo or in-vitro applications, where the gate and channel are exposed to the liquid medium and are connected to an encapsulated circuit based on traditional electronics. The same idea can be transferred to other applications, such as the recording of biopotentials  \cite{khodagholy2015neurogrid}. 

\subsection{Device-to-Device Crosstalk}
\label{sec:coupling}
While sensing or stimulation can be carried out using individual devices or interfaces, many other applications, such as digital computation, require a more complex circuitry based on multiple components. Already simple logic gates need a number of transistors. The integration of multiple OECTs in the same electrolytic medium, where each component is in electrical contact due to its ionic conductivity, is very challenging  \cite{romele2020multiscale}. This is an odd scenario when compared to traditional circuit design based on solid-state components such as MOSFETs, where each device is (ideally) uncoupled and independent.
Occasionally, demonstrations of OECT circuitry are carried out by casting individual drops of electrolyte in such a way that each channel is in contact only with its "parent" gate. Such an approach cannot be imagined to be translated into a commercial product nor real-life applications. Moreover, experiments like these can result in poor consistency due to the quick evaporation rate of small water drops, and consequential difficulty in maintaining a constant electrolyte concentration. All these problems can be  circumnavigated at once by employing solid and patternable electrolytes, ion-conductive gels, or crosslinked molten salts. 
The use of patternable ion-conductors allowed for high-performance OECTs and power-efficient ICs  \cite{weissbach2022photopatternable,andersson2019all}. Solid OECTs have been successfully demonstrated for a variety of applications and allowed to leverage the peculiar properties of the OMIECs and OECTs for traditional circuitry. 
However, while OECTs based on solid-state electrolytes are being explored as a good alternative to other solid-state circuit elements, electrochemical transistors operated in liquid electrolytes are arguably the best example of circuit elements that can be employed in a biological wet environment for computing, as they combine very large capacitance, good electronic-to-ionic transduction, stability in physiological conditions, and biocompatibility.
Under this "global condition", individual gating is difficult to achieve and significant crosstalk between devices is to be expected. Crosstalk is a well-known effect present in any IC under different forms and names (near-field coupling, capacitive coupling, leakage currents, inductive coupling, common-impedance coupling, etc.).  Crosstalk typically carries a negative connotation as it refers to a mechanism where signals influence each other in undesired ways. Clearly, capacitive and inductive effects are used in many applications e.g., wireless communications, but typically the "action at distance" occurs between two well defined components rather than with the whole circuit indiscriminately, often referred to as parasitic currents.  Many parameters affect the magnitude of the crosstalk, from the amplitude and frequency of the signals to the material and geometry of the circuit. Therefore, careful circuit design must foresee and address such an effect. In OECT circuitry designed to work in electrolytic environments, crosstalk cannot be avoided even in simple designs. This coupling is difficult to model due to the nonlinear electrical characteristics of OECTs and because of the voltage and concentration-dependence of the electrical property of the liquid. Therefore, in order to employ multiple OECTs and OECT-based devices in a circuit, such dynamics must be understood and harnessed.
Two regimes can be distinguished, static and dynamic, to analyze the coupling between devices, based on the timescale the act on. The characteristic charging time of a channel $\tau$ i.e., the time needed to switch from the on to the off state and vice-versa is proportional to the product of the capacitance of the channel (hence its volume), and the resistance of the gating medium  \cite{rivnay2015high}.
It is important to note that, although the circuit is often modelled with linear components (resistors and capacitors), electrolyte and electrolyte/solid interfaces are generally nonlinear and the linear approximations may break down close to current saturation state both in the on- and off-state.
If the voltages applied to the devices are time-dependent and vary on a time scale much longer than $\tau$, a static regime can be assumed.

\subsection{Static regime}
In this case, Eq. \ref{eq:oect} is a good approximation, although the contributions of all the voltages applied in the global environments must be summed up.

 For example, the current flowing in a channel under the influence of two (identical, i.e., same size) gates, as sketched in Fig. \ref{fig:panel2}e is
\begin{equation}	
	I_d =G_0 (1- \frac{V_{g,1}+V_{g,2} - \frac{V_d}{2}}{V_p})V_d.
	\label{eq:twogates}
\end{equation}
Importantly, in the static regime, the distance between channel and gate does not matter as long as the electrolyte provides good ionic conduction. Therefore, a channel will be switched on and off almost identically by any gate immersed in solution  \cite{cindyspaper}.
Eq. \ref{eq:twogates} breaks down if the gates are not identical. Different size brings about different capacitance which must be summed in parallel. Moreover, by engineering the gate material or its surface functionalization, the channel may respond differently depending on the electrolyte composition. Such a technique can be used, for example, to build ion-specific sensors. Another simple case occurs when two identical channels are biased with $V_{d,1}$ and $V_{d,2}$ (Fig. \ref{fig:panel2}f), each one gating the other, although the gating effect is weaker due to its space dependency over the channel  \cite{cucchi2021reservoir}. The currents $I_{d,1}$ and $I_{d,2}$ flowing through the two channels obey

\begin{equation}
	I_{d,1} = G_0 (1 -\frac{V_{d,2} - V_{d,1} }{2V_p})V_{d,1}
	\label{eq:parallel1}
\end{equation}
and, symmetrically,
\begin{equation}
	I_{d,2} = G_0 (1 -\frac{V_{d,1} - V_{d,2} }{2V_p})V_{d,2}.
\label{eq:parallel2}
\end{equation}	
Putting together Eqs. \ref{eq:oect}, \ref{eq:twogates}, and \ref{eq:parallel1}, the current flowing in the channel in a configurations such as the one in Fig. \ref{fig:panel2}g is
\begin{equation}	
	I_{d,1} =G_0 (1- \frac{V_{g,1}+V_{g,2} + \frac{V_{d,2}}{2} - \frac{V_d,1}{2}}{V_p})V_{d,1}.
	\label{eq:twooects}
\end{equation}

The unwanted crosstalk influences heavily the current-voltage characteristics of each channel, hampering the use of such devices for purposes where static behavior is expected, such as digital electronics.
 Only simplistic  (one-input-one-output) circuit components can be built and be functional under such conditions, for example complementary inverters  and sensors  \cite{romele2020multiscale}. Logic gates that require two inputs, fundamental in digital electronics, are already challenging due to the fact that both gates affect all the channels indiscriminately, strongly limiting the usefulness of OECTs for logic in-liquido. These obstacles, however, can be worked-around operating in the dynamic regime.\\

\subsection{Dynamic Regime}
\label{sec:globalgating}
\begin{figure}	
	\includegraphics[width=\linewidth]{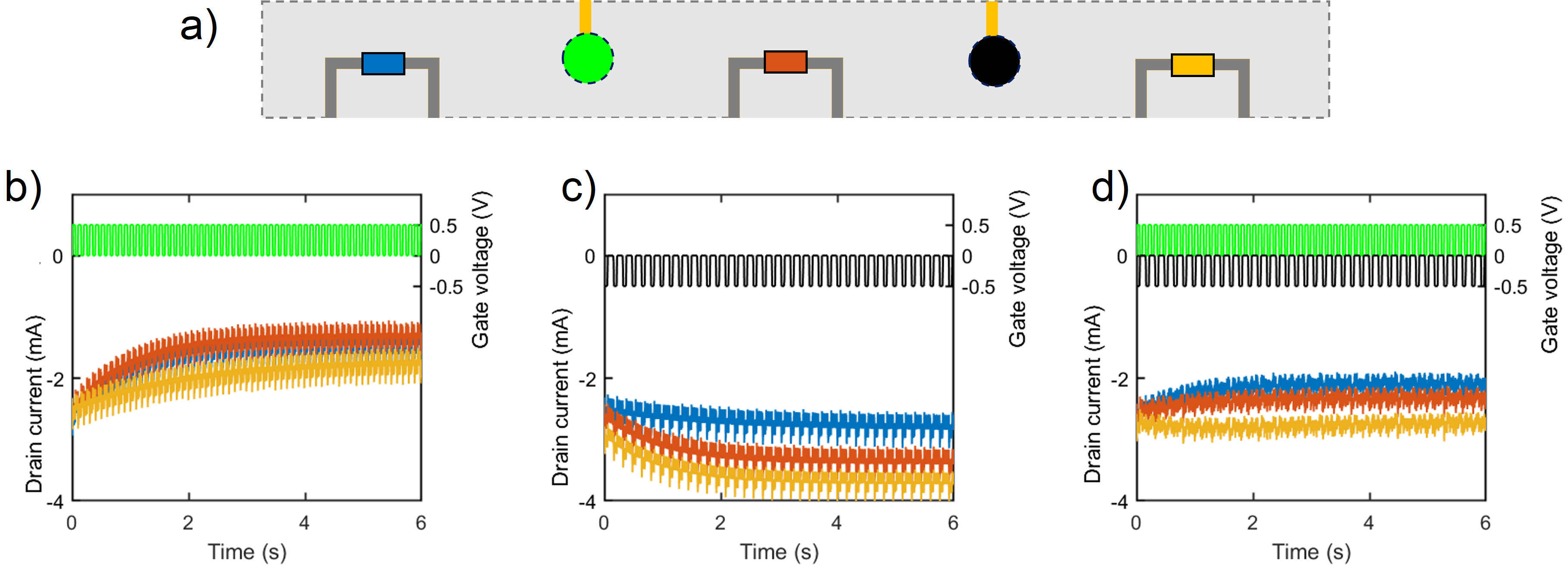}
	\caption{Example of a two-gate-three-channel configuration. a) Sketch of the physical circuit with PEDOT:PSS channels of 100x200 $\mu$m biased at $V_d=500$ mV and gates of 500x500 $\mu$m in water:NaCl 100 mM. The color code of each component is maintained in the plots below for data visualization. b-c) Drain currents of the three channels under the effect of one gate pulsing and d) with both gates pulsing simultaneously.}
	\label{fig:pulses}
\end{figure}

When time-dependent voltages are applied and varying on a time scale comparable to, or lower than $\tau$ (typically ranging in tens of millisecond), the ionic (dis)charging time must be taken into account (assuming that the response of the electronic circuit of the OECT is orders of magnitude faster than the ionic circuit). Using a simple in-series RC circuit, for a pulsed gate voltage of amplitude $V_g$ starting at time t=0, the current in the channel obeys
 \begin{equation}
 	 	I_d(t)=G_0  (1- (\frac{V_g - \frac{V_d}{2}}{V_p}) (1-\exp(-\frac{t}{\tau})))V_d 
 	\label{eq:timedep}
 \end{equation} 
for a normally-on OECT  \cite{bernards2007steady}. Therefore, $I_d(t)$ depends on the distance between circuit components and adds a degree of freedom with respect to Eqs. \ref{eq:twogates}, \ref{eq:parallel1}, and \ref{eq:parallel2}. Again, this is a linear approximation of an otherwise nonlinear element. A deeper analysis of the transient time of OECTs can be found in Refs.  \cite{faria2017transient, paudel2022transient}.
The dynamic regime offers more freedom and opportunities for integration of multiple devices within the same electrolyte. For example, while in a simple two-gate-two-channel configuration as in Fig. \ref{fig:panel2}g, finding the combination of gate voltages that fix the current of the channels to two specific values can be impossible due to the crosstalk, the situation is different if pulses are applied.
The optimization of the pulse duration and amplitude in relation to the distance between each gate-channel pair allows to set the current of the channels to the desired values. Naturally, it will be a time-varying current swinging between two levels due to the time-dependent nature of the gate signals. Therefore one should refer to their average current or a temporal integration. This is a necessary step towards brain-inspired and physical computation.
This could be harnessed in a design in which pulsed gate signals are used to modulate the current in neighboring channels without affecting farther channels. An example is reported in Fig. \ref{fig:pulses}, where  it is displayed how the resistance of three channels is affected by individual gates, and how their effect can be superimposed to modulate each channel's (average) conductance.

The nonlinear coupling of all the potentials applied within the electrolyte offers new routes to explore different criteria for circuit design. The idea of using the transient time for electronics and computation goes hand-in-hand with recent trends in the development of new computational paradigms that exploit dynamical systems and transient effects. Moreover, biologically-inspired applications, such as spiking neurons and growing networks, are more easy to deploy using dynamic effects.

This approach is in stark contrast with traditional digital electronics (where the transients are negligible and crosstalks are minimized) and it is often referred to as "in-materio computing". Hence, in analogy with this term, in-materio computing in wet environments is named "in-liquido computing". 

\begin{figure}	
	\includegraphics[width=\linewidth]{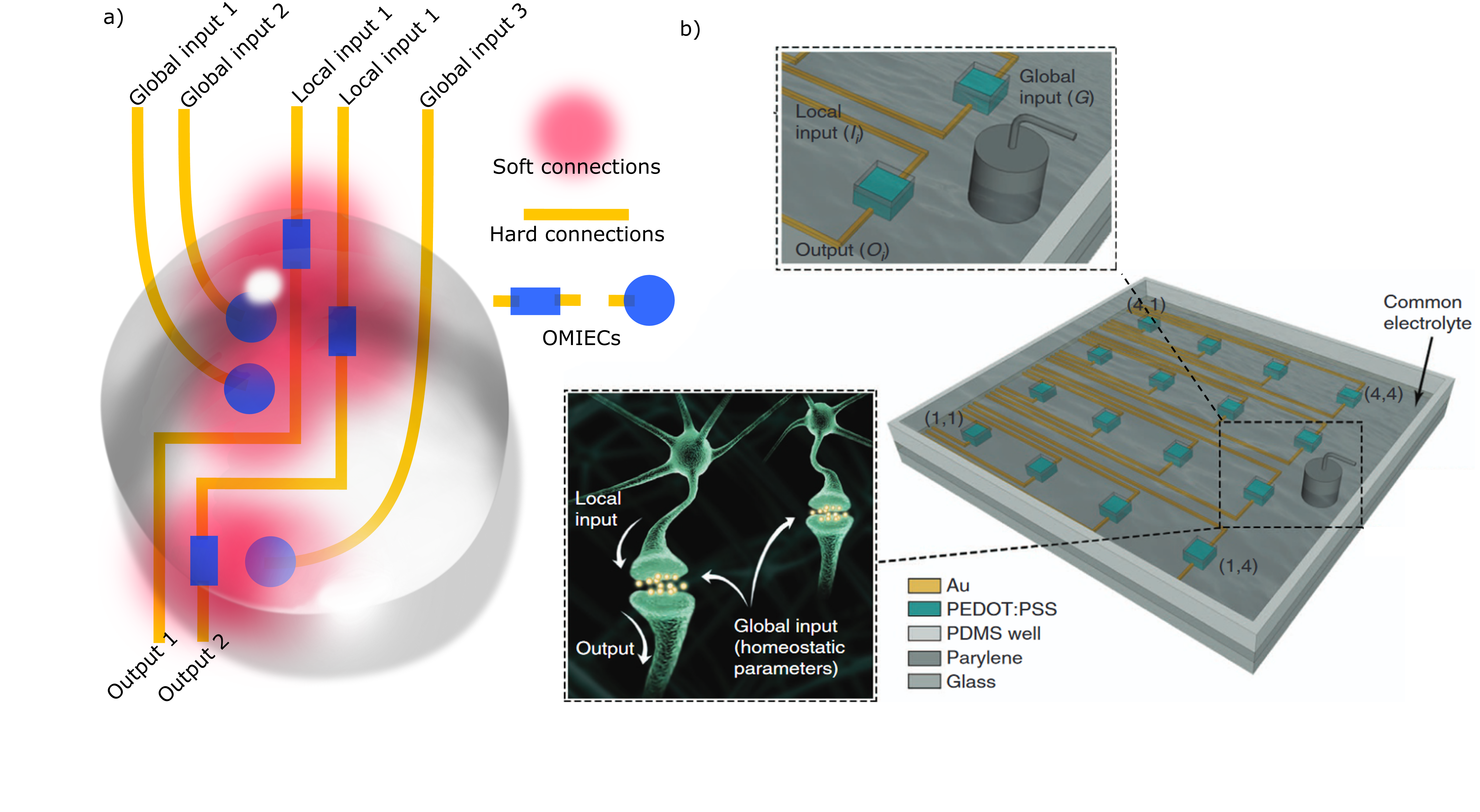}
	\caption{a) Top-view on an exemplary circuit made of OMIEC-based channels and gates with highlighted hard connections (the physical organic film bridging metal tracks) and soft connections (all the voltage-mediated interactions between each component). This approach was first explored by Gkoupidenis \textit{et al.}  \cite{gkoupidenis2017neuromorphic} as shown in panel b): here, a single gate exerts a global input on a multitude of channels that, however, interact with each other. In the inset, a pair of neurons interact with global and local inputs. 
	}
		\label{fig:soft_hard_connections}

\end{figure}
\subsection{Collective Behavior: Beyond digital}
\label{sec:collective}
While circuit elements are generally designed to respond to a defined set of inputs, which are fed by means of such metallic connections, biological "electrical" circuitry and wetware evolved to work very differently: not only local inputs drive the operation of cells, nerves and synapses, but a global input too (temperature, viscosity, chemical composition, etc.) affects their response and behavior. At the same time, the environment has a direct effect on the network, acting on it to maintain it stable and avoid drifting in operation ranges too close to hyperactivity or quiescence  \cite{turrigiano2004homeostatic}.

While the wetware is a completely different paradigm compared to our software/hardware design, plenty of examples in nature show how even simple organisms can carry out complex tasks (communication, navigation, playing, fighting, recognition) with much smaller energy requirements than our best robotic systems.
The above-mentioned influence on a number of global and local parameters is mediated by the chemical composition of the environment and on the activity of neighboring "circuit elements" (as it is the case for close neurons firing). The modification of these global parameters, referred to as homeostatic environment,  allows for a prompt and collective change of the  behavior of the whole system. Moreover, it is used to affect the magnitude of the crosstalk between near yet unconnected neurons, and maintains the activity of the network at the optimal level of excitability without drifting into a range of hyperactivity or quiescence  \cite{turrigiano2004homeostatic}.
These connections, which are not physically built, rather mediated by the environment, can be regarded as soft connections, a term firstly coined in Ref.  \cite{gkoupidenis2017neuromorphic}.
Although the manipulation of the soft connections is a powerful tool to influence the behavior of a circuit without the need of constant formation and disruption of physical connections, the nervous system is also capable of physical rewiring: neurons can grow dendrites and synapses, they can change size and even die in order to establish the optimal connectivity within the network. These are physical (or hard) connections, and they are a good analogy to the metallic tracks in traditional circuits (of course, metallic lines do not show synaptic plasticity). A sketch of soft and hard connections is depicted in Fig. \ref{fig:soft_hard_connections}.\\

With this scenario in mind, whereby both soft and hard connections play a key role, OMIECs come in extremely handy. 
Firstly, the OMIEC can be used as metallic track for the physical connections. Secondly, OECTs can be employed as active elements (transistors, artificial synapses, spiking neurons, sensors) whose response will depend necessarily both on the local input and on the global inputs (gating). Lastly, OMIECs can be grown in the same electrolytic environment where (and under similar condition at which) they operate. These observations naturally points toward the realization of evolvable and reconfigurable electronics, a topic that we address in Section \ref{sec:evolvable}. 

In principle, simple circuitry could operate correctly even in global connectivity by clever design of the components and of the pulse scheme used to drive them. The idea of employing global connectivity in OECTs was pioneered by Gkoupidenis \textit{et al.}, both in a many-gate-single-channel   \cite{gkoupidenis2016orientation,gkoupidenis2016orientation2} and in single-gate-many-channel architectures  \cite{gkoupidenis2017neuromorphic}. This early work stems from the necessity of finding a spatio-temporal correlation between input and output signals in a global environment (homeostatic conditions) to design brain-inspired electronics  \cite{gkoupidenis2015neuromorphic}. 
 For instance,  orientation selectivity, a key mechanism in the lower level of the visual system (i.e., the visual cortex), can be reproduced with a single-channel OECT addressed with a grid of gates. Moreover, by forcing global electrochemical oscillations in an array of such devices, the activity of those devices can be synchronized  \cite{koutsouras2019functional}. Phase-dependent coupling between the global signal the local device activity is induced by such global oscillations. This coupling induces a functional type of connectivity between neuromorphic devices in analogy with the brain oscillations (also known as brain waves) that synchronize distant neural populations. This is in strong analogy with the brain, where despite the significant noise and stochastic neurons spiking, the neural networks have robust firing properties observable in stable and global oscillation patterns. 
 
 In a similar way, coincidence detection, a mechanism important for the auditory system, but also similar to the spatio-temporal integration of incoming inputs of the neuron soma, can be implemented with OECT arrays  \cite{gkoupidenis2017neuromorphic}. Building upon these concepts, multi-gate arrays can be employed for ion-mediated multiplexing  \cite{koutsouras2021iontronic} and spatio-temporal input correlation  \cite{qian2017multi}. In this configuration, the spatio-temporal response of multi-gate arrays can be used for \textit{blind} \textit{local} ionic multiplexing. \textit{Blind} means that there is no need for address assignment at the inputs (in contrast to classic multiplexing), and \textit{local} means that multiplexing is a result of the spatio-temporal properties of the device without depending on classic peripheral circuity. The above mentioned examples show the flexibility and complexity in ionic communication that can be achieved with the incorporation of global electrolytes in neuromorphic architectures. 
 These works exploits the transient time originating from the (dis)charging of an OECT channel for spatio-temporal discrimination. In order to implement such architectures into useful electronics, the digital I/O paradigm must be replaced with a time-continuous spike-like implementation, similar to how neurons "decide" whether to fire or not.
 
\section{Evolvable, self-organized, and random networks}
\label{sec:evolvable}
 \begin{figure}	
	\includegraphics[width=\linewidth]{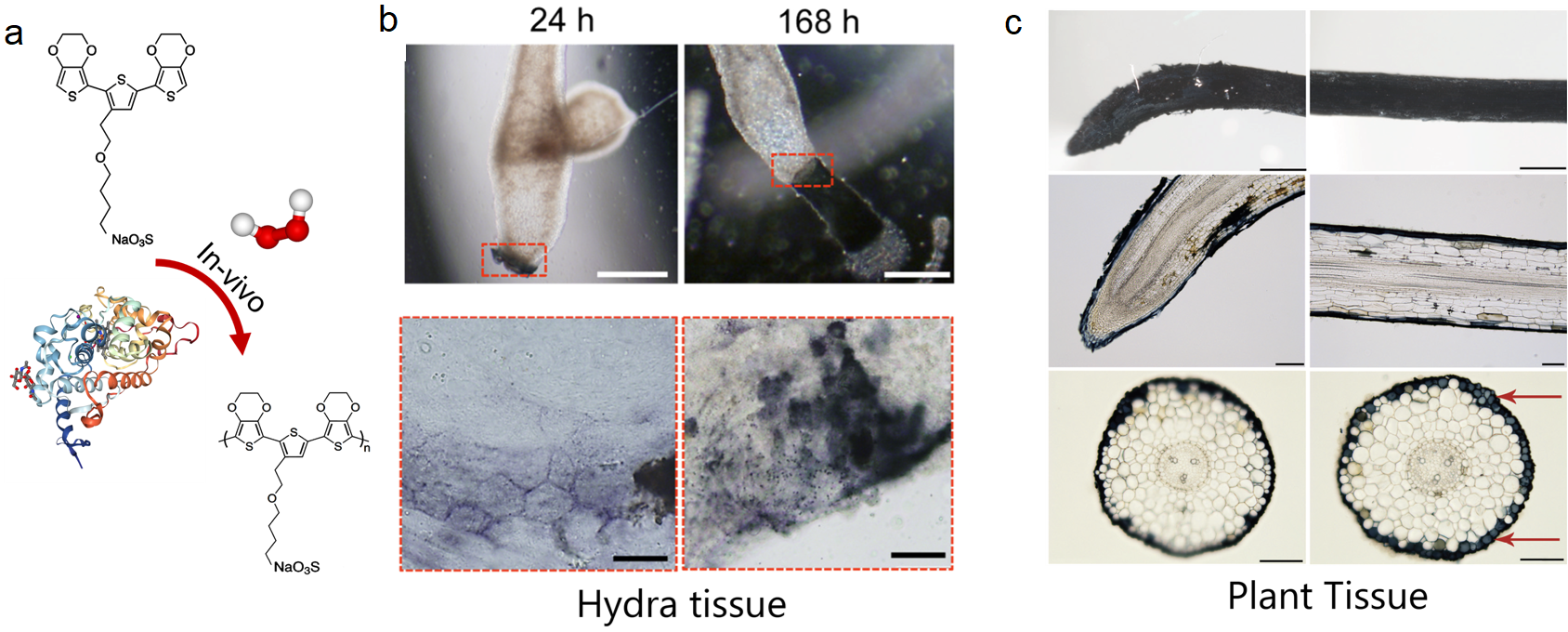}
	\caption{a) ETE-S polymerizes in-vivo due to endogenous peroxidase enzymes and hydrogen peroxide.  b) Formation of OMIEC in the basal disc area of hydra polyps treated with ETE-S after 24 h and 168 h of incubation. Scale bars 500 $\mu$m (upper panel), 50 $\mu$m lower panel. c) Bean plants roots functionalized with ETE-S resulting in a OMIEC coating on the roots epidermis, upper: plane view (scale bar 500 $\mu$m), middle: longitudinal cross sections (scale bar 100 $\mu$m), and lower: transversal cross sections (scale bar 100 $\mu$m).}
		\label{fig:danis}

\end{figure}
Utilizing new materials and physical systems to build implants may also be accompanied by unconventional fabrication and deployment techniques that differ from the traditional micro and nanofabrication of ICs. Achieving good, long-lasting, coupling between the fabricated system and the targeted tissue is challenging and at the same time a fundamental step to overcome to achieve intimate interface. 
As introduced in Sec. \ref{sec:collective}, OMIECs have the peculiarity that the electrolytic environment can be both a catalyst for their polymerization and the medium for their operation as transistors. Therefore, a circuit of OMIECs can be modified during operation allowing for a reconfiguration and organization of the circuitry. 
If the electrical pathways evolve under specific circumstances occurring during their operation, the final circuit will shape itself in a way that depends on the history of its operation. In an analogous way, the nervous system restlessly rewires its neuronal circuitry based on the body's activity and its internal and external stimuli. As such, evolvable OECTs and OMIECs have the potential to produce brain-inspired circuitry.


Gerasimov et al. took the first prototypical steps towards unsupervised learning and reconfigurable circuitry using OECTs by demonstrating in-operando oxidative polymerization of OMIECs and so producing  evolvable organic electrochemical transistors  \cite{gerasimov2019evolvable}. They further coupled the trigger of the growth with other physical, environmental, and analog processes within the circuit itself  \cite{gerasimov2021biomimetic}. The technique involves the dissolution in the electrolyte of the precursor monomer of the OMIEC and an ion capable of doping it. If a potential larger than the oxidation potential of the monomer is applied, the electropolymerization starts.
Electropolymerization is typically obtained by applying a constant voltage or current against a counter electrode. In a similar fashion, an AC signal can be employed  \cite{cucchi2021directed,ciccone2022growth}. The dendritic and filament-like morphology resulting from the AC electropolymerization, rather than closed-films as in the DC case, allows for guided and directional connection growth. Such method can be leveraged for the production of blueprint-less networks for in-vitro and in-vivo sensing platforms and (neural) networks as it will be discussed in Sec. \ref{sec:computation}  \cite{cucchi2021reservoir}.
 The electropolymerization process can be controlled only by a certain extent leading to poor device homogeneity. Therefore, these devices are more suitable for computational approaches where randomness and complexity are desirable rather than, for instance, crossbars for matrix-vector multiplications  \cite{van2018organic,van2017non}.
The great effort devoted to develop and optimize the growth of evolvable OECTs did not solve a key issue: a controllable "depolymerization" process is still missing. A depression mechanism to decrease the weight of connections it is mandatory for reliable ANN training. Two ways to decrease a connection's conductivity are available: dedoping the film with a gate electrode, or waiting for its degradation. The former introduces the problem of crosstalk with other connections hence lowering the accuracy of the network, the latter is a passive, slow and uncontrolled process. Nonetheless, the potential of evolvable and electropolymerized organic conductive films for bioelectronics is further exemplified by examples of in-vivo formation of the OMIECs   \cite{wilks2011vivo,murbach2018situ}.  
In one example, the active site of an implanted neural probe is formed in-situ via electropolymerization of dispersed EDOT monomers, thereby decreasing the interfacial impedance and improving the signal to noise ratio of the recordings. Moreover, Koizumi \textit{et al.} demonstrated that the growth can be triggered "wirelessly" employing an external electrical field  \cite{koizumi2016electropolymerization}. \\
These techniques may allow for an accurate growth of these polymers in-vivo to target specific tissues or cells, however the driving current of the polymerization may be detrimental. An alternative to this issue comes from the use of biological substrates as a scaffold and trigger for the polymerization of OMIECs, opening the doorway to interfacing bioelectronics to the biocatalytic machinery of biological organism, and intimate coupling to biological wetware. In this framework, EDOT-thiophene based derivatives were synthetized as water soluble oligomers with the aim of developing self-doped OMIECs suitable for biofabrication  \cite{mantione2020thiophene}.   A recent example  involves the invertebrate hydra as an animal model to polymerize 4-(2-(2,5-bis(2,3-dihydrothieno[3,4-b] [1,4]dioxin-5-yl)thiophen-3-yl)ethoxy)butane-1-sulfonate (ETE-S), as shown in Fig. 5a. The conducting polymer was localized in tissues that expressed peroxidase activity, but it was also found in the adhesive substance secreted by the animal that it used to attach on surfaces while underwater  \cite{tommasini2022seamless} (Fig. 5b).

The approach for self-organization and polymerization of oligomers triggered by the biochemical environments was first demonstrated in plants  \cite{stavrinidou2017invivo}.
ETE-S polymerized in the vascular tissue of rose cuttings without any physical or chemical stimuli forming conducting wires with conductivity of 10 S/cm and specific capacitance of 20 F/g. In a later work it was demonstrated that peroxidase enzymes, present in the plant cell wall were responsible for the polymerization reaction (Fig 5a)  \cite{dufil2020enzyme} while the plant tissue is acting as a template for the polymer that organizes in a favorable manner with pronounced $\pi - \pi$ stacking  \cite{parker2022biohybrid}. When intact plants were “watered” with ETE-S solution, an integrated conducting root network was formed and, not only it did not hamper the plant development, but it led to the growth of a more complex root system, as the plant was adapting to the new hybrid state (Fig. 5c). As for the roots conductivity, it remained stable  for over four weeks even though the roots kept growing showcasing the effective functionalization.  

These works represent examples of harnessing the biocatalytic machinery of living organisms for the fabrication and seamless integration of electroactive interfaces in-vivo. Both in plants and in hydra animal model the integrated OMIEC is biocompatible without any signs of toxicity or negatively impacting the biological function.
In plants, the tissue integrated OMIECs in combination with the natural electrolytic environment were used to demonstrate the possibility of utilizing the in-vivo organized OMIECs in devices \cite{stavrinidou2015electronic}. Such components can be used to communicate and better leverage the intrinsic computational capabilities of plant's physiological dynamics  \cite{pieters2022leveraging}. For example, in a hydra model, the ETE-S was polymerizing during the foot regeneration at the amputation site forming a conducting layer simultaneously with the newly forming tissue. Although these devices required addressing with external electrodes and did not have a record performance, they open the pathway for a new paradigm on how technological components are integrated in living tissues. Further understanding of how to utilize the complex biochemical environment of living organisms, not only for organizing functional components, but also to enable bi-directional communication between the two may result in autonomous systems that self-regulate the biotic and abiotic interaction and evolve over time. 

\section{Computation and Machine Learning Using Electrochemical Networks}
\label{sec:computation}
Looking at a biological neural network, it is not trivial to clear state which is the fundamental unit carrying and processing information. The most numerous in the biological, as well as in artificial neural networks is the synapse. Artificial synapses are devices whose resistance can be tuned at wish. These must then be implemented in large networks and be used in computational tasks.
While a large pool of devices for electrolyte-compatible artificial synapses has been demonstrated, only a handful of examples pointed towards the implementation at the network level  \cite{van2017non,fuller2019parallel}, and rarely in liquid environment.
Akai-Kasaya \textit{et al.}  \cite{akai2020evolving} employed the electropolymerization technique mentioned in Sec. \ref{sec:evolvable} to grow conductive fibers made of PEDOT:PSS between gold electrodes. The directionality of the reaction can be controlled through the amplitude and frequency of the triggering AC field and ensures good control over the process. They demonstrated feed-forward neural networks where the weight of each connection is controlled by the reaction time.

\begin{figure}
	\includegraphics[width=\linewidth]{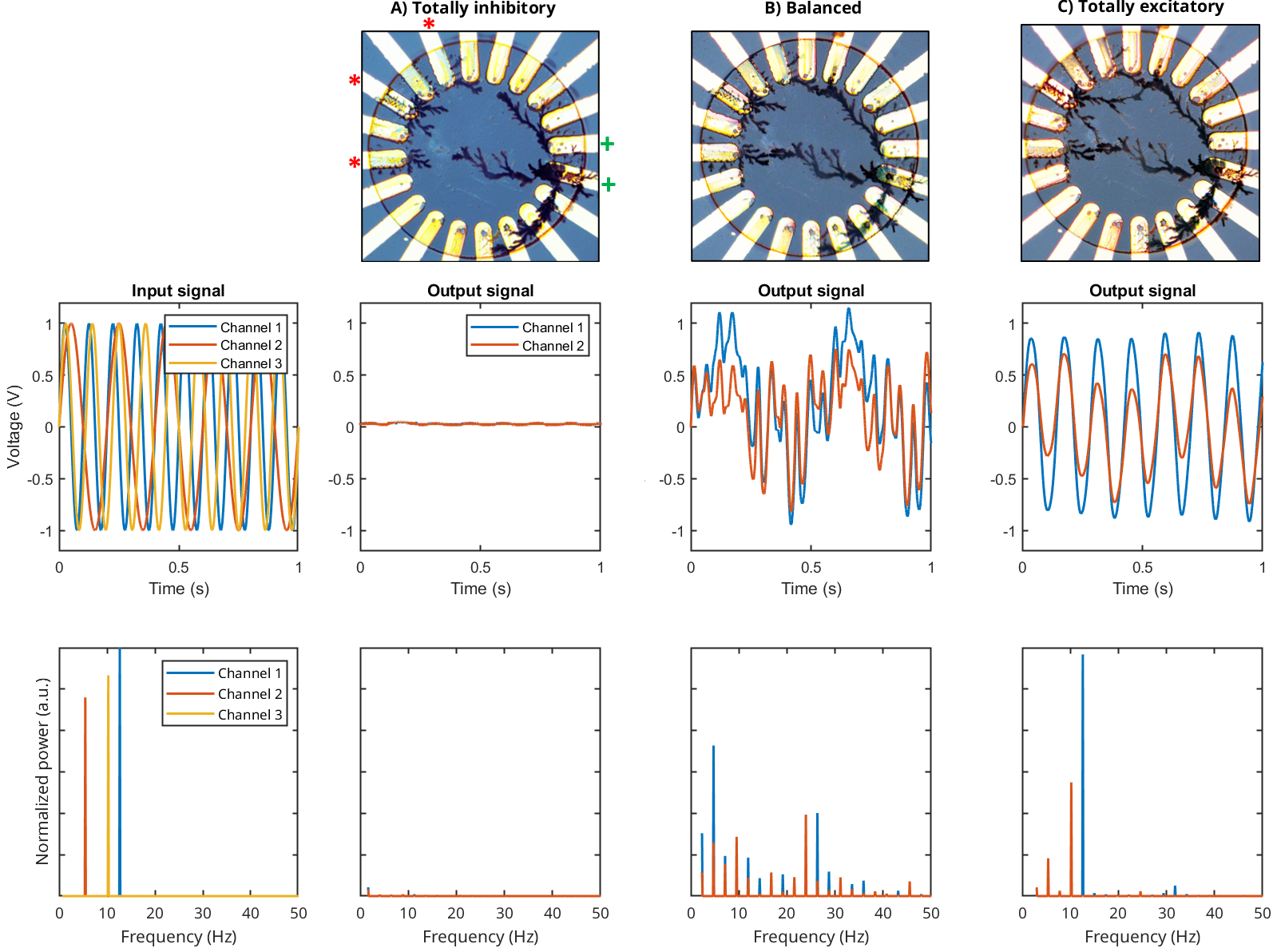}\vspace{1cm}
		\includegraphics[width=\linewidth]{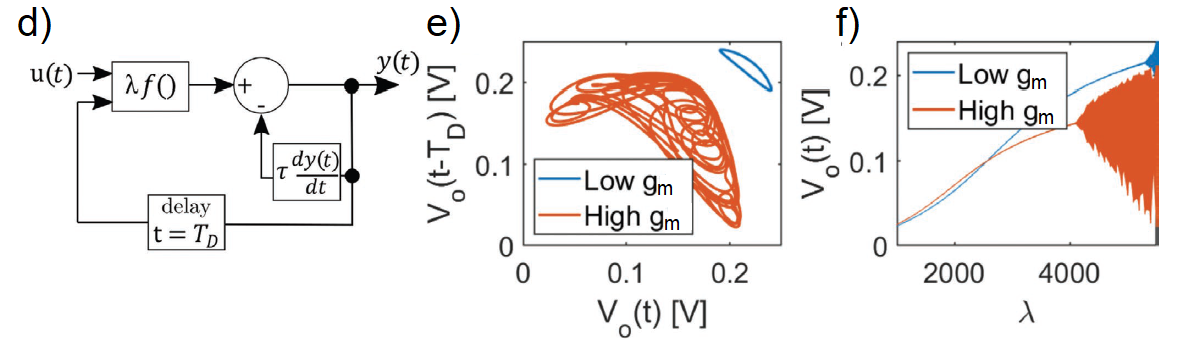}
	\caption{Behavior of electrochemical networks: a) a network based on ionic currents i.e. no hard connections between input and output layers result in poor nonlinearity as shown by the output curves in time and frequency domains. b) a balance between soft and hard connections is ideal for information processing and c) a fully connected network behaves mostly linearly.
	d) Employment of a delayed-feedback loop in an electrochemical network  \cite{petrauskas2021nonlinear} may transform a nonlinear network into a chaotic one, hence lending much more complexity to the physical system (where $\lambda$ is an amplification factor). e) The input voltage is plotted vs the output voltage for OECTs with different transconductance ($g_m$), and f) its relative bifurcation plot.} 
	\label{fig:design}
\end{figure}

Precise synaptic connectivity (i.e., fine-tuning of the edges) is a prerequisite for a functional neural circuit, be it a biological, physical, or algorithm-based. However, brain rewiring is a developmental growth mechanism that is not only characterized by accuracy, but also fault-tolerance, flexibility and robustness. Similar to other growth processes, cellular interactions are restricted in space and time, especially considering a synaptic density in humans of 1x10$^9$ synapses/mm$^3$  \cite{huttenlocher1979synaptic}. As a result, noise is not negligible in neuronal firing and an important source of stochasticity, and when subjected to the same inputs, neurons and neural networks can respond in different ways. 
These observations raised the doubt: \textit{can a network incapable of tuning precisely each of its units be functional?} Such question can be brought to its extreme: \textit{can a completely untrained network still perform useful computation?}\\
The quest intrigued computer scientists and the first models of computation performed by random and fixed networks were already proposed in the 80's. More recently, the method was introduced in a general machine learning framework: by coupling a random, fixed, untrained network with a perceptron-like output layer (which is the only part of the network trained via linear regression), the method came to be known as reservoir computing (RC)  \cite{maass2002real,jaeger2004harnessing}. RC networks can be used effectively for information processing while unloading the burden of the time and energy-consuming  training phase of the network that, by definition, is random and does not need any design blueprint nor adjustment of its weights. By driving input signals into a random, large, fixed recurrent neural network, and thereby inducing in each neuron within this "reservoir" a nonlinear response, one can obtained a desired output signal by a trainable linear combination of all of these output signals  \cite{Tanaka2019,cucchi2022hands}. The random network/layers mediate a nonlinear mapping of the input layer onto the higher-dimensional space of the output layer. This is the so-called kernel method.
How RC compares to traditional ML is outside the scope of this work, but it undoubtedly stands out as a promising contender for physical neural networks. This method can leverage the countless nonlinear properties that can be found in material systems, as well as circumnavigate the necessity of precise training and low device-to-device variability. Indeed, RC has been demonstrated on a widely diverse plethora of systems, from traditional silicon-based transistors to more exotic examples employing nonlinear photonic or magnetic effects  \cite{Tanaka2019,cucchi2022hands}.
Among these, chemical reservoirs (i.e., reservoirs that use nonlinear dynamics coming from chemical reactions) can be exploited to produce computational systems for specific applications in liquid environments and may directly interface with the biochemical machinery of our body. As chemical dynamics are limited in speed by the reaction and diffusion rates, chemical reservoirs do not stand out for speed and are not suitable for applications where low-latency is required. However, when dealing with biological systems, the speed becomes secondary and other properties, such as cytocompatibility, long-term stability, and interaction with the environments are prioritized.  
Very recently, examples of chemical RC that capitalized on redox reactions have been proposed, showing excellent performance and great potential for biosensing as well as for machine learning in-liquido. For examples, Kan \textit{et al.}  exploited the nonlinear spatial and temporal dynamics of an electrolytic solution containing polyoxometalate  \cite{kan2022physical}. Such a compound can undergo multiple oxidation and reduction states, producing Faradaic reactions with the metallic electrode upon the application of a voltage. The authors showed that an electrolytic solution alone can be used a random network for periodic time-series predictions. 
The usefulness of the electrolyte and electrochemical reactions can be better leveraged in RC if the ionic dynamics are converted into electronic signals using OMIECs and OECTs. A first proof-of-concept demonstration was devised by Pecqueur \textit{et al.}  \cite{pecqueur2018neuromorphic}, who used a gate as "source" time-dependent signal and an array of 16 OECT channels continuously biased. The gate signal modulates the current in the channels. Without any device-to-device variability, each channel should respond identically to the gate signals. However, intrinsic imperfections in the morphology of the polymeric materials create inherent variability. 
Key in this work is the demonstration that 1) device-to-device variability, an ubiquitous factor in organic electronics can be harnessed as a feature and that 2) ionic-to-electronic transduction using OECTs can be used for machine learning. They showed that such a system can be used to distinguish different time-dependent signals, like square waves from triangular waves. In this work,  the proposed reservoir is still "consciously-designed" and requires precise microfabrication of the transistor channels.\\
Cucchi \textit{et al.} built upon this idea and produced reservoirs made of electropolymerized OECTs. The electropolymerization offers several advantages such as inexpensive lithography-free fabrication as well as dendritic topology  \cite{cucchi2021reservoir}. The latter emerges as an important feature when several fibers are operated in global gate conditions. The voltage drops along a fiber between the two metal connections, making up for a hard connection, or, as named in the article, an excitatory connection. However, if a fiber branches out and terminates in the electrolyte, its potential will be constant as its resistance is much lower than the one of the electrolyte, constituting an inhibitory (soft) connection.  
The negligible voltage drop along inhibitory fibers enhances the gating strength of such branches, whose effect on neighboring and farther fibers builds up during the charging transient time. This grants rich nonlinear behaviors  \cite{petrauskas2021nonlinear} as shown in Figure \ref{fig:design}. 
Nonlinear, synchronized and even chaotic patterns arise as a consequence of synergistic interactions between each fiber segments: each fiber, driven by local inputs, modifies the behavior of others operating in the same environment. At the same time, its action is modified by the others. The consequence of this mutual modification, operating over time, is the emergence of a stable pattern.
As it often happens in emergent systems, the precise design of its components is irrelevant for the final appearance of complexity, but a few requirements for integration must be obeyed. Fig. \ref{fig:design} highlight the importance of a balance between hard and soft connections. Fig.\ref{fig:design}a shows the poor nonlinear transfer response of a networks that are completely formed by fibers connecting the input to the output layer (totality of excitatory fibers), as well as formed by fibers that terminate within the solution (totality of inhibitory fibers). A balance between the two is needed for a good nonlinear transformation, although a quantitative analysis is missing. 
 Clever design (although still inherently random or semi-random)  of the networks allows for remarkable results in machine learning tasks, such as flower recognition using the iris dataset. The networks worked much better using time-varying signals, highlighting the usefulness of the ionic dynamics and the nonlinear charging transient of the fibers.
When a delay-line is added to the network, therefore implementing a single-node reservoir computing architecture, the nonlinearities of the system are extremized into a chaotic behavior  \cite{petrauskas2021nonlinear} as shown in Figs. \ref{fig:design}d-f. This way, complex tasks such as time-series predictions and heartbeat recognition can be carried out with excellent accuracy, 96\% and 88\%, respectively   \cite{cucchi2021reservoir}. 
Along these lines, Usami \textit{et al.} covered a set of circular electrodes with a film of sulfonated polyaniline  \cite{usami2021materio}. The film was then immersed in an electrolytic solution. This way, no patterning at all is needed, and the nonlinear dynamics are mediated by dishomogeneities of the film as well as by the spatial distribution of the injected and recorded signals. Such method was successfully employed for time-dependent classification such as spoken digits with an accuracy of 70\%.\\

These approaches demonstrate that complex computation can be attained in-liquido at low energy cost exploiting the dynamical properties of a given material system. This can be a useful extension of the toolbox of emergent multimodal implantable electronics  \cite{huang2022actively}  thanks to the inherent low-power consumption due to the analog signal processing, the biocompatibility of OMIECs, and the fact that traditional electronics suffers from electrolyte-induced instability and require long-term hermetic encapsulations. 
\section{Closed-loop bioelectronics}
This perspective focused on the use of electrochemical transistors for the design and fabrication of intelligent and implantable computational units. We did not, purposefully, dig into the possibility of using OMIECs as sensors, although sensors 
are key components for closed-loop biointerfaces as they produce the input signals that must be computed and classified. 
As observed in Sec. \ref{sec:computation}, the sensing and computing mechanisms do not need to be separated units because of the intimate correlation between electrolytic environment and OMIEC response that ultimately mediate the nonlinear response of the network. Therefore, the only missing element, is the reaction of the system to a specific input. As introduced in the first section and in Fig. 1, the potential of adaptive and smart circuitry for implantable devices lie in the possibility of combining the sensing process with efficient information processing that happens locally on the substrate: the result of such computation must result in action. Similarly to the biochemical feedback loops that make up for our physiological processes, human-made devices must be able to interact with the surroundings not only by collecting information but also by acting upon it.
For this to happen, the output of the computational unit i.e. the network, has to be connected to an element capable of acting on the sensed/computed information i.e. an actuator. An optimally trained network driven in current or voltage will activate only one neuron of the output layer i.e. will classify correctly the inputs and result in a clear answer. The current or voltage at this neuron, then, can be amplified and used to drive an actuator. This could be, for example, a pulse generator, an optical signal, or a drug release system.
The choice of the system at this stage is arbitrary. However, OMIECs stand out also for this task. Firstly, metallic interfaces used to stimulate or record signals can be covered with conductive polymers in order to lower dramatically their impedance, hence reducing the need of bulky power sources. Moreover, clever chemical design can lead to the incorporation of charged biomolecules within the polymeric matrix that, upon the application of a voltage, can be released in the surroundings  \cite{castagnola2017multilayer,kleber2019electrochemically}. 
The actuator will inevitably modify the environment sensed by the network which must be then robust under these changes. In this way, after an initial training, the closed-loop system can work unsupervised and autonomously.

\section{Outlook}
When showcasing single-device performances and projecting their systems for bioelectronics, authors should be solicited to pay attention on the final aim of circuit integration, and the properties of the device must be analyzed accordingly e.g., in biological environments or phosphate-buffered saline solution, where device-to-device crosstalk may occur.
In parallel, new approaches can be exploited in order to harness connectivity and global oscillations. These approaches are of particular value in neuromorphic computing, machine learning, and more in general biologically-inspired information processing, where chemical signals orchestrate complex dynamics in a common electrolytic environment. Moreover, neural networks can be obtained by clever exploitation of the nonlinear dynamics arising from the crosstalk. In this scenario, device-to-device variability and random designs are desirable. This route could lead to energy-efficiency and human-compatible hardware neural networks for on-chip (or edge) computing.\\
To better face these challenges, numerical circuit simulations capable of predicting the response of multiple devices globally connected must be developed in order to run quick circuit modeling and optimization and accelerate the circuit design. At the current state, such quest is particularly difficult considering that even at a single-device level, simulations and experiments for OECTs and OMIECs are often in poor agreement. Moreover, although the electrolyte can be simulated using dense meshes of capacitors and resistors, these tools and user interfaces are not optimized for such tasks. Finally, "nonidealities" such as chemical mechanisms and diffusive dynamics cannot be overlooked - in fact, they may be key to the rich temporal dynamics which seem to be essential for power efficient on-chip computing.
\section{Conclusions}
The design, fabrication, and modeling of electronics that can operate in liquid environments has been drawing increasing attention because conventional electronics seem to be insufficient for applications such as autonomous and power-efficient implants capable of information processing.
Mixed ionic-electronics conductors and OECTs emerged as ideal candidates for such a quest, although a blueprint to realize intelligent circuitry in liquid seems to be missing.
 This Perspective dealt with the challenges, workarounds and novel ideas concerning the implementation of complex circuitry based on OECTs, and can be extended to other devices that can be gated through electrolytes. We have highlighted the recent trends, from the modeling standpoint to the implementation. Finally, recent applications of machine learning have been unpacked revealing great potential for implantable $in-liquido$ computation using OMIECs and electrochemical transistors.

\bibliography{template.bbl}
\section*{Acknowledgments}
ES and DP acknowledge funding from the Swedish Research Council (VR-2017-04910) and the Swedish Foundation For Strategic Research (FFL18-0101)
\section*{Conflict of Interest}
All authors declare no conflict of interest
\end{document}